\\
Title: Is space-time curved?
Authors: Z. Junhao and C. Xiang
Comments: 13 pages with 2 figures
Report-on:
Subj-class: General Physics
\\

Whether the space-time is curved or not? The experimental criterion to judge this point are: (1) The experimental results of red shift, deflective angle of light and perihelion shift of planet in essence are favorable to the gravitational theory based on flat space time, but unfavorable to the general relativity. (2) In Gravity Probe-B experiment, if the experimental value of angular acceleration of gyroscope of orbital (geodetic) effect is

$$\overline{\left(\frac{d\mathbf{w}_L^{(S)}}{dt}\right)} = \frac{GM_0}{c^2 r^3}(\mathbf{r} \times \mathbf{v}) \times \mathbf{w}_L^{(S)}$$

and the experimental value of angular acceleration of the earth rotation (frame-dragging) effect projected in the equatorial plane is

$$\overline{\left(\frac{d\mathbf{w}_S^{(S)}}{dt}\right)}_{x_1=0} = \frac{3GM_0 R^2 \Omega}{10 c^2 r^3} w_{S1}^{(S)} \mathbf{j}$$

then the space time is flat. If the experimental values of these two angular accelerations of gyroscope are

$$\overline{\left(\frac{d\mathbf{w}_L^{(G)}}{dt}\right)} = \frac{3GM_0}{2c^2 r^3}(\mathbf{r} \times \mathbf{v}) \times \mathbf{w}_L^{(G)}$$

$$\overline{\left(\frac{d\mathbf{w}_S^{(S)}}{dt}\right)}_{x_1=0} = \frac{GM_0 R^2 \Omega}{5 c^2 r^3}(-w_{S2}^{(G)} \mathbf{i} + w_{S1}^{(G)} \mathbf{j})$$

respectively, then the space time is curved.
\\

Is space-time curved? If we want to answer this question, we must clarify the experimental criterion to judge whether the space-time is flat. However Newton gravity is not a formula that it satisfies the demand of covariance of special relativity. Therefore its results cannot be used to represent the character of flat space-time. If there is a new gravitational theory, that it set up based on the flat space-time and satisfies the covariance of special relativity. Then we can deduce the experimental criterion of flat space-time from this theory.

In our previous work[1–6], we set up a new theory–special relativistic gravitational theory. Now let us compare the mathematical construction of special relativistic gravitational theory with general relativity, both are set up based on different space-time. Therefore we can learn how to judge whether space-time flat.

**(1) Difference of mathematical construction between two gravitational theories**

As discussed in our previous studies, the first hypothesis of special relativistic gravitational theory is the equivalence of inertial mass and gravitational mass. Based upon this hypothesis one would have the 4-dimensional gravitational force as follows

$$K_n = m_0 H_{nsr} U_s U_r \qquad (1)$$

where $U_s$ is the 4-dimensional velocity of the exerted body, $H_{nsr}$ is the strength of gravitational field. According to the special relativistic relation

$$K_n U_n = 0 \qquad (2)$$

the 3-dimensional expression of gravitational force is

$$\mathbf{F} = m(\mathbf{E} + \frac{1}{c}\mathbf{u} \times \mathbf{B} + \frac{1}{c}\mathbf{u} \cdot \wp + \frac{1}{c^2}\mathbf{u} \times \Re \cdot \mathbf{u}) \qquad (3)$$

where $\mathbf{u}$ is the 3-dimensional velocity of the exerted body, and

$$E_i = -c^2 H_{i44}$$

$$B_i = \frac{ic^2}{2}(H_{jk4} + H_{j4k} - H_{kj4} - H_{k4j})$$

$$P_{ij} = \frac{ic^2}{2}(H_{ij4} + H_{i4j} + H_{ji4} + H_{j4i})$$

$$R_{il} = \frac{c^2}{3}(H_{jkl} + H_{jlk} - H_{kjl} - H_{klj}) \qquad (4)$$

It is important that the fourth term of Eq.(3) is perpendicular to $\mathbf{u}$ from Eq.(2). The mathematical form of Eq. (1) is the same as the equation of general relativity

$$\frac{dU^{n(G)}}{dt} = -\Gamma^n_{sr} U^{s(G)} U^{r(G)} \qquad (5)$$

where symbols $(S)$, $(G)$ are used to distinguish the corresponding quantities in special relativity and general relativity. Therefore, the question focus on the **difference** between $-\Gamma^n_{sr}$ and $H_{nsr}$.

From Eq. (2) and the symmetry of $H_{nsr}$ for $s$ and $r$, we can obtain

$$H_{nsr} = \frac{\partial A_{sr}}{\partial x_n} - \frac{1}{2}\frac{\partial A_{nr}}{\partial x_s} - \frac{1}{2}\frac{\partial A_{ns}}{\partial x_r} \qquad (6)$$

The first term of Eq. (3) is independent of 3-dimensional velocity (except $m$ factor), so this term is the Newton gravity. From Newton gravitational potential and the covariance of gravitational field equation under Lorentz transformation, we get that the only possible form of the gravitational field equation is

$$\frac{\partial}{\partial x_z}\frac{\partial}{\partial x_z}(A_{mn} + l\delta_{mn}A_{aa}) = -\frac{(1+4l)4\pi G\rho_0}{(1+3l)c^4}U_m U_n \tag{7}$$

where $l$ is a constant. If we choose that $l = -1/2$, then the second hypothesis of the special relativistic gravitational theory is that, the equation of the gravitational field has the form

$$\frac{\partial}{\partial x_z}\frac{\partial}{\partial x_z}(A_{mn} - \frac{1}{2}\delta_{mn}A_{aa}) = -\frac{8\pi G\rho_0}{c^4}U_m U_n \tag{8}$$

Newton potential is the summary of the classical gravitational observation, which is only related to weak field. Therefore, this equation is merely the linear approximation of the field equation.

In the weak field limit, the metric tensor of general relativity is in the form

$$g_{mn} = \eta_{mn} + h_{mn} \tag{9}$$

where $\eta_{mn}$ is the constant metric tensor

$$\eta_{mn} = \begin{pmatrix} 1 & 0 & 0 & 0 \\ 0 & 1 & 0 & 0 \\ 0 & 0 & 1 & 0 \\ 0 & 0 & 0 & -1 \end{pmatrix} \tag{10}$$

$h_{mn}$ and their derivatives are small quantities of the first order. In this case, the field equation of general relativity is simplified as

$$\frac{\partial}{\partial x_a}\frac{\partial}{\partial x_a}h_{mn} = \frac{-8\pi G\rho_0}{c^2}[\eta_{mn} + \frac{2}{c^2}U_m^{(G)}U_n^{(G)}] \tag{11}$$

Because the right side of Eq. (11) contains a $G$ factor, $U_m^{(G)}$ may only take the zeroth order approximation. Under such circumstance, we have $U_m^{(G)} \approx \eta(\mathbf{u}, -c)$. In the linear approximation, the equation of $A_{mn}$ is analogous to the equation of $h_{mn}$. Then we have

$$\frac{1}{2}h_{mn} = \begin{pmatrix} A_{ij} & iA_{i4} \\ iA_{4j} & -A_{44} \end{pmatrix} \tag{12}$$

However the physical meaning of $A_{mn}$ is different from $h_{mn}$. According to general relativity, the relation between field strength and potential is

$$-\Gamma^n_{sr} = \frac{1}{2}\eta^{na}(h_{sr,a} - h_{ra,s} - h_{as,r}) \tag{13}$$

Substituting Eq. (12) into Eq. (13) and comparing with Eq. (6), we have

$$-\Gamma^l_{sr} = \begin{pmatrix} H_{lij} + N_{lij} & i(H_{li4} + N_{li4}) \\ i(H_{l4j} + N_{l4j}) & -(H_{l44} + N_{l44}) \end{pmatrix} \tag{14}$$

$$-\Gamma^4_{sr} = -i\begin{pmatrix} H_{4ij} + N_{4ij} & i(H_{4i4} + N_{4i4}) \\ i(H_{44j} + N_{44j}) & -(H_{444} + N_{444}) \end{pmatrix} \tag{15}$$

which

$$N_{nsr} = -\frac{1}{2}\left(\frac{\partial A_{ns}}{\partial x_r} + \frac{\partial A_{nr}}{\partial x_s}\right) \tag{16}$$

$N_{nsr}$ represents the **fundamental difference** between the two gravitational theories.

Let us rewrite Eq. (1) as

$$\frac{dU_4^{(S)}}{dt} = c^{-2} E_j U_4^{(S)} U_j^{(S)} + ic^{-2} P_{ij} U_i^{(S)} U_j^{(S)} \tag{17}$$

$$\frac{dU_i^{(S)}}{dt} = c^{-2}[-E_i (U_4^{(S)})^2 - i(U_j^{(S)} B_k - U_k^{(S)} B_j + P_{il} U_l^{(S)}) U_4^{(S)} + (U_j^{(S)} R_{kl} - U_k^{(S)} R_{jl}) U_l^{(S)}] \tag{18}$$

Using Eqs. (14), (15) and (4), we can rewrite Eq. (5) as

$$\frac{dU^{4(G)}}{dt} = iN_{444}(U^{4(G)})^2 + (c^{-2} E_j + 2N_{44j}) U^{j(G)} U^{4(G)} + (c^{-2} P_{ij} - iN_{4ij}) U^{i(G)} U^{j(G)} \tag{19}$$

$$\frac{dU^{i(G)}}{dt} = c^{-2}\{(E_i + E_i')(U^{4(G)})^2 + [U^{j(G)}(B_k + B_k') - U^{k(G)}(B_j + B_j') +$$
$$+ (P_{il} + P_{il}')U^{l(G)}]U^{4(G)} + (U^{j(G)} R_{kl} - U^{k(G)} R_{jl}) U^{l(G)}\} + N_{ijl} U^{j(G)} U^{l(G)} \tag{20}$$

which

$$E_i' = -c^2 N_{i44} = c^2 \frac{\partial A_{i4}}{\partial x_4}$$

$$B_k' = ic^2 (N_{ij4} - N_{ji4}) = \frac{1}{3} B_k$$

$$P_{ij}' = ic^2 (N_{ij4} + N_{ji4}) = -P_{ij} - 2ic^2 \frac{\partial A_{ij}}{\partial x_4} \tag{21}$$

If all of the $N_{nsr}$ terms are zero, the dynamic equation of general relativity will become the corresponding equation in special relativistic theory. Therefore, the main focus of the problem becomes **whether we can verify that all the $N_{nsr}$ terms are zero from experiment.**

### (2) The motion of a body in the gravitational field due to a static sphere $M_0$

From Eq.(8), we obtain that the gravitational potential due to the static sphere $M_0$ is

$$A_{mn} = \frac{GM_0}{c^2 r} \begin{pmatrix} 1 & 0 & 0 & 0 \\ 0 & 1 & 0 & 0 \\ 0 & 0 & 1 & 0 \\ 0 & 0 & 0 & -1 \end{pmatrix} \qquad r > R \qquad (22)$$

which $R$ is the radius of sphere. Then we get

$$E_i = \frac{-GM_0 x_i}{r^3}, \qquad B_i = 0, \qquad \wp_{ij} = 0,$$

$$R_{il} = \frac{GM_0}{r^3} \begin{pmatrix} 0 & x_3 & -x_2 \\ -x_3 & 0 & x_1 \\ x_2 & -x_1 & 0 \end{pmatrix} \qquad r > R \qquad (23)$$

Therefore Eq. (3) reduces to

$$\mathbf{F} = m(\mathbf{E} + \frac{1}{c^2} \mathbf{u} \times \mathfrak{R} \cdot \mathbf{u}) \qquad (3')$$

The first term of this formula is Newton gravity. The second term is additional gravity, it is perpendicular to $\mathbf{u}$ from Eq. (2). So it is not to do work. In this case, Eqs. (17), (18) become

$$\frac{dU_4^{(S)}}{dt} = c^{-2} E_j U_4^{(S)} U_j^{(S)} \qquad (24)$$

$$\frac{dU_i^{(S)}}{dt} = c^{-2}[-E_i (U_4^{(S)})^2 + (U_j^{(S)} R_{kl} - U_k^{(S)} R_{jl}) U_l^{(S)}] \qquad (25)$$

The energy $w$ of body is proportional to $U_4^{(S)}$, so from Eq. (24), we obtain the energy integral

$$w = w_0 \exp\left[\frac{GM_0}{c^2}\left(\frac{1}{r} - \frac{1}{r_0}\right)\right] \approx w_0 \left[1 + \frac{GM_0}{c^2}\left(\frac{1}{r} - \frac{1}{r_0}\right)\right] \qquad (26)$$

From Eq. (25), we get that the angular momentum integral is

$$L = L_0 \exp\left[-\frac{GM_0}{c^2}\left(\frac{1}{r} - \frac{1}{r_0}\right)\right] \approx L_0 \left[1 - \frac{GM_0}{c^2}\left(\frac{1}{r} - \frac{1}{r_0}\right)\right] \qquad (27)$$

There are two fundamental integrals if a body moves in the gravitational field due to a static sphere $M_0$. According Eqs. (26), (27), we obtain the orbital equation of a body is

$$\left(\frac{d\mathbf{r}}{d\mathbf{j}}\right)^2 = -\left[1-(8-2k^2)\frac{G^2M_0^2}{h^2c^2}\right]\mathbf{r}^2 + (4-2k^2)\frac{GM_0}{h^2}\mathbf{r} + \frac{c^2(1-k^2)}{h^2} \quad (28)$$

which $\mathbf{r} = 1/r$, $k$ and $h$ are two constants. The solution of Eq. (28) has the form

$$\mathbf{r} = A[1 + e\cos(\mathbf{mj})] \quad (29)$$

According Eqs. (28), (29), we can find

$$\mathbf{m} \approx 1 - \frac{3GM_0 A}{c^2} = 1 - \frac{3GM_0}{c^2 a(1-e^2)} \quad (30)$$

for the motion of planet. So the perihelion shift of the planet is

$$\Delta \mathbf{j} \cong 2\mathbf{p}(1-\mathbf{m}) = \frac{6\mathbf{p}\, GM_0}{c^2 a(1-e^2)} \quad (31)$$

This is identical with the result of experimental observation.

Eqs. (26), (27) are still correct for the motion of a photon. If a photon passes from the Sun ($r_0 = R$) to the earth ($r \approx \infty$), the variance of energy is

$$|\Delta w| = \frac{GM_0 w_0}{c^2 R} \quad (32)$$

From $w = h\mathbf{n}$, we obtain

$$\frac{\Delta \mathbf{n}}{\mathbf{n}} = \frac{GM_0}{c^2 R} \quad (33)$$

This is also in agreement with the experimental observation. The red shift experiment directly proves that the additional gravity is not to do work, the conclusion of flat space-time is correct.

If we take $k = 0$ in Eq. (28), it is an equation for the motion of photon, then we have

$$\left(\frac{d\mathbf{r}}{d\mathbf{j}}\right)^2 = -\left(1 - \frac{8G^2M_0^2}{c^2 h^2}\right)\mathbf{r}^2 + \frac{4GM_0}{h^2}\mathbf{r} + \frac{c^2}{h^2} \quad (34)$$

The solution of the above equation can be written in the form of Eq. (29). Substituting it into Eq. (33), we may obtain

$$e = \frac{ch}{2GM_0} \quad (35)$$

Finally we get that the deflection angle of light is

$$\mathbf{d} = \frac{4GM_0}{c^2 R} \quad (36)$$

It is identical with the experimental value of deflective angle of light.

These facts express that: (i) the special relativistic gravitational theory **identically**

**describes** the motion of a body in the static gravitational field. The results of three classical relativistic gravitational experiments cannot be used to prove that the space-time is not flat. (ii) The experimental value of $N_{nsr}$ is

$$N_{nsr} = 0 \qquad (37)$$

Now let us discuss the next question. Do these experimental results **truly** prove that the space-time is curve? When a body moves in the gravitational field due to static sphere, Eqs. (19), (20) are reduced to

$$\frac{dU^{4(G)}}{dt} = \left(\frac{1}{c^2}E_j + 2N_{44j}\right)U^{j(G)}U^{4(G)} \qquad (38)$$

$$\frac{dU^{i(G)}}{dt} = \frac{1}{c^2}E_i(U^{4(G)})^2 + \frac{1}{c^2}\left(U^{j(G)}R_{kl} - U^{k(G)}R_{jl}\right)U^{l(G)} + N_{ijl}U^{j(G)}U^{l(G)} \qquad (39)$$

According to general relativity, $N_{44j}$ and $N_{ijl}$ are non-zero, whereas the rest of the $N_{nsr}$ terms are zero, where $N_{44j}$ has the following expression:

$$2N_{44j} = \frac{1}{c^2}E_j = \frac{-GM_0 x_j}{c^2 r^3} \qquad (40)$$

Similarly, we can obtain the energy integral of general relativity from Eq. (38),

$$w^{(G)} = w_0^{(G)} \exp\left[\int_{r_0}^{r}\left(c^{-2}E_j + 2N_{44j}\right)dx^j\right]$$

$$\approx w_0^{(G)}\left[1 + \int_{r_0}^{r}\left(c^{-2}E_j + 2N_{44j}\right)dx^j\right] = w_0\left[1 + \frac{2GM_0}{c^2}\left(\frac{1}{r} - \frac{1}{r_0}\right)\right] \qquad (41)$$

so we have

$$w - w_0 = \frac{2GM_0 w_0}{c^2}\left(\frac{1}{r} - \frac{1}{r_0}\right) \qquad (42)$$

The factor 2 is extremely important. We rewrite the general relativity equation in forms of Eqs. (19), (38). The purpose is only to clearly trace the source of the factor 2. In Eq. (41) $w_0^{(G)}\int_{r_0}^{r}c^{-2}E_j dx^j$ represents the variance of energy originates from Newton gravity, and $w_0^{(G)}\int_{r_0}^{r}2N_{44j}dx^j$ represents the variance of energy from the non-Newton terms, which is predicted by general relativity. Eq. (40) indicates that both terms have the equal contribution which resulting in the factor 2 in Eq. (42).

From the first order approximation of Schwerzchild exact solution, we also obtain

$$\Delta w = \frac{2GM_0 w}{c^2}\left(\frac{1}{r} - \frac{1}{r_0}\right)$$

$$\Delta L = 0 \tag{43}$$

In general, people believe that three classical relativistic experiments support (original) general relativity. This only means that: the perihelion shift of the planet and the deflective angle of light deduced from Eqs. (42), (43), and the frequency red shift deduced from metrical tensor are identical with the observations. However these are incomplete because from the relation $w = h\mathbf{n}$ and the experimental value of frequency red shift, the variance of energy is

$$|\Delta w| = \frac{GM_0 w}{c^2 R} \tag{44}$$

Eq. (42) deduced from (original) general relativity is in contradiction to the experimental value by a factor 2. So the exact statement is that (original) general relativity only obtains the support of **two and a half experiments**. Caused by this fundamental contradiction, many people are trying to come up with new theories or corrections. All these new theories will be tested by the new experiment, especially by experiments like Gravity Probe-B.

The rational deduction should be:
- The variance energy in red shift experiment points out that the factor 2 in Eq. (42) is not correct. This fact means that non-Newton term does not contribute to the variance energy. The experimental value of $N^{(e)}_{44j}$ must be zero.
- There is no reason to believe that Eq. (42) is still correct in other experiments, e.g. the deflection of light. General relativity explains the perihelion shift of the planet and the deflective angle of light, but its validity is doubtful because the correctness of Eq. (42) needs to be justified.
- If we delete the factor 2, then Eq. (42) becomes Eq. (26). In this case, if we want to obtain the experimental values of perihelion shift of planet and deflection angle of light in the static source problem. The only way is to combine Eq. (26) with Eq. (27). It means that the experimental value of $N^{(e)}_{ijl}$ must be zero. So all experimental values $N^{(e)}_{nsr}$ are zero in the static source problem.

Are the above interpretations correct? They can be tested by GP-B experiment. If the experimental value of orbital (geodetic) effect precession rate is 4.4 sec/yr then the above inferences is correct.

**(3) The precession of gyroscope in the gravitational field due to the moving source**

When the source is moving, the components of field due to this source are $\mathbf{B} \neq 0$, $\wp \neq 0$. Both gravitational theories predict that the gravity contains the terms, which are linear relative to 3-dimensional velocity of the exerted body (except $m$ factor). The second and third terms of Eq. (3) are proportional to $u/c$. So in the case of low velocity, the effect of these two terms is much smaller than that of the Newton term. The best way to test whether these two terms exist in experiment is to put a rotating gyroscope in the gravitational field and to measure its precession rate. In this case, the Newton term does not contribute to the precession rate of the gyroscope, therefore people can measure the size of the term linear relative to 3-dimensional velocity. Of cause, the precession rate is very small. If people want to obtain reliable results, the measurement must be very accurate which is extremely difficult.

If gyroscope is a well-distributed sphere, from Eq. (3), we can obtain

$$\frac{d\mathbf{w}}{dt} = \frac{1}{2c}[(\mathbf{w} \times \mathbf{B}) - (\wp \cdot \mathbf{w}) + \mathbf{w}\, Sp(\wp)] \tag{45}$$

If $\mathbf{B} \to \mathbf{B} + \mathbf{B}'$, $\wp \to \wp + \wp'$, we shall obtain the relative equation in general relativity.

NASA and Stanford University are planning to "testing Einstein with orbiting gyroscopes — gravity probe B". In this experiment, four spinning gyroscopes will be set up in an orbiting satellite around the earth. When the satellite is in the orbital motion, from the satellite instantaneous rest-frame, the earth is moving around the satellite. Also the earth revolves around its axis of rotation. The spinning gyroscope is used to probe $\mathbf{B}$ and $\wp$ components of gravitational field due to these two types motion of the earth. To distinguish two types of field, let us use the symbol $\mathbf{B}_L, \wp_L$ to represent the component of field due to the motion of the earth around the satellite, and use $d\mathbf{w}_L/dt$ to represent the angular acceleration caused by these components of field and simply call it as orbital effect. Symbolize $\mathbf{B}_S$, $\wp_S$ as the components of field due to the earth revolve around its rotation axis, and $d\mathbf{w}_S/dt$ to represent the angular acceleration caused by these components of field, and simply call it as earth rotation effect.

### (3-1) The orbital (geodetic) effect

In the earth frame, the center of the earth is static. The gravitational potential due to the static sphere is Eq. (22). Using the Lorentz transformation, we can get the gravitational potential observed on the satellite instantaneous static frame and finally obtain:

$$\mathbf{B}_L = \frac{3GM_0}{cr^3}(\mathbf{v} \times \mathbf{r}) \tag{46}$$

$$\wp_L = \frac{-GM_0}{cr^3}[\mathbf{vr} + \mathbf{rv} - (\mathbf{v} \cdot \mathbf{r})\Im] \tag{47}$$

$$\mathbf{B}_L' = \frac{1}{3}\mathbf{B}_L \tag{48}$$

$$\wp_L' = -\wp_L + \frac{2GM_0 v x_1}{cr^3}\Im \tag{49}$$

where $\mathfrak{I}$ is the $3\times 3$ unit tensor, $\mathbf{r}$ is the distance from the earths center to the satellite, $\mathbf{v}$ is the velocity of satellite observed on the earth static frame. Because the satellite frame is an accelerating frame, it is needed to add the Thomas precession term into Eq. (45), therefore we have

$$\frac{d\mathbf{w}_L^{(S)}}{dt} = \frac{1}{2c}[\mathbf{w}_L^{(S)} \times \mathbf{B}_L - \mathbf{w}_L^{(S)} \cdot \wp_L + \mathbf{w}_L^{(S)} Sp(\wp)] - \frac{GM_0 (\mathbf{r}\times\mathbf{v}) \times \mathbf{w}_L^{(S)}}{2c^2 r^3} \qquad (50)$$

Because the period of satellite revolution is shorter than the period of the gyroscope precession, it is appropriate to take the average value of strength field to substitute the instantaneous value. So we have

$$\overline{\left(\frac{d\mathbf{w}_L^{(S)}}{dt}\right)} = \frac{GM_0}{c^2 r^3}(\mathbf{r}\times\mathbf{v}) \times \mathbf{w}_L^{(S)} \qquad (51)$$

In general relativity, we must consider the contribution of $N_{nsr}$, therefore we have

$$\overline{\left(\frac{d\mathbf{w}_L^{(G)}}{dt}\right)} = \frac{3GM_0}{2c^2 r^3}(\mathbf{r}\times\mathbf{v}) \times \mathbf{w}_L^{(G)} \qquad (52)$$

This formula is the same as that given by Schiff,L.I. (1960) which determines the geodetic precession rate predicted by general relativity. GP-B group pointed out that the geodetic precession rate is 6.6 sec/yr then the orbital effect precession rate predicted by special relativistic gravitational theory will be 4.4 sec/yr

In the above discussion, we take two fundamental hypotheses of special relativistic gravitational theory as the premise to deduce the precession rate of orbital effect. However, there are other ways to calculate the precession rate. Taking the results of three classical relativistic experiments as the premise, we can find out the formula of gravity and all component of strength field due to a static sphere. Using the Lorentz's transformation, we can obtain the component of strength field due to a moving sphere. Then we can calculate the precession rate of orbital (geodetic) effect. The final result is the same as Eq. (51). So it is extremely likely that people obtain the value of precession rate of orbital effect predicated by special relativistic gravitational theory from GP-B experiment. If the experimental value of precession rate of orbital effect approaches 4.4 sec/yr this fact will prove that our inference for three classical relativistic experiments is correct.

### (3-2)The earth's rotation (frame-dragging) effect

From Eq. (8), we may obtain the potential and strength of field due to the earth rotation

$$\mathbf{B}_S = \frac{3GM_0 R^2}{5cr^5}[3\mathbf{r}\times(\mathbf{W}\times\mathbf{r}) - 2r^2 \mathbf{W}]$$

$$= \frac{3GM_0 R^2 \Omega}{5cr^5}(-3x_1 x_3, -3x_2 x_3, x_1^2 + x_2^2 - 2x_3^2) \qquad (53)$$

$$\wp_S = \frac{3GM_0 R^2 \Omega}{5cr^5} \begin{pmatrix} -2x_1 x_2 & x_1^2 - x_2^2 & -x_2 x_3 \\ x_1^2 - x_2^2 & 2x_1 x_2 & x_1 x_3 \\ -x_2 x_3 & x_1 x_3 & 0 \end{pmatrix} \quad (54)$$

$$\mathbf{B}_S' = \frac{1}{3}\mathbf{B}_S \quad (55)$$

$$\wp_S' = -\wp_S \quad (56)$$

which $\Omega$ is the angular velocity of the earth rotation, $ox_3$ is the earth axis, $ox_2 x_3$ is the satellite orbital plane. Therefore we get

$$\begin{aligned}\frac{d\mathbf{w}_S^{(S)}}{dt} = &\frac{3GM_0 R^2}{10c^2 r^5}\{\mathbf{w}_S^{(S)} \times [3\mathbf{r}\times(\mathbf{W}\times\mathbf{r}) - 2r^2 \mathbf{W}] \\ &- \Omega[(-2w_{S1}^{(S)} x_1 x_2 + w_{S2}^{(S)}(x_1^2 - x_2^2) - w_{S3}^{(S)} x_2 x_3)\mathbf{i} \\ &+ (w_{S1}^{(S)}(x_1^2 - x_2^2) + 2w_{S2}^{(S)} x_1 x_2 + w_{S3}^{(S)} x_1 x_3)\mathbf{j} \\ &+ (-w_{S1}^{(S)} x_2 x_3 + w_{S2}^{(S)} x_1 x_3)\mathbf{k}]\}\end{aligned} \quad (57)$$

$$\frac{d\mathbf{w}_S^{(G)}}{dt} = \frac{2GM_0 R^2}{5c^2 r^5}\mathbf{w}_S^{(G)} \times [3\mathbf{r}\times(\mathbf{W}\times\mathbf{r}) - 2r^2 \mathbf{W}] \quad (58)$$

Eq. (58) is the same as that given by Schiff, L.I. (1960). The frame-dragging precession rate predicated by general relativity is deduced from this formula.

Because the period of satellite revolution is shorter than the period of gyroscope precession, it is appropriate to take average values, then we get

$$\overline{\left(\frac{d\mathbf{w}_S^{(S)}}{dt}\right)}_{x_1=0} = \frac{3GM_0 R^2 \Omega}{10c^2 r^3} w_{S1}^{(S)}\mathbf{j} \quad (59)$$

$$\overline{\left(\frac{d\mathbf{w}_S^{(G)}}{dt}\right)}_{x_1=0} = \frac{GM_0 R^2 \Omega}{5c^2 r^3}(-w_{S2}^{(G)}\mathbf{i} + w_{S1}^{(G)}\mathbf{j}) \quad (60)$$

The contribution of $\wp$ term exists in special relativistic gravitational theory, which is not the case in general relativity, so the law of precession is difference in these two theories.

Define

$$\mathbf{e}_j = \frac{1}{\sqrt{w_{S1}^2 + w_{S2}^2}}(-w_{S2}\mathbf{i} + w_{S1}\mathbf{j}) \quad (61)$$

and

$$\dot{\mathbf{y}}_S = \frac{1}{\mathbf{w}_S}\left(\frac{d\mathbf{w}_S}{dt}\cdot\mathbf{e}_j\right) \quad (62)$$

therefore we get

$$\dot{\mathbf{y}}_S^{(S)} = A\sin q \cos^2 j \quad (63)$$

$$\dot{\psi}_S^{(G)} = \frac{2}{3} A \sin q \qquad (64)$$

In addition

$$A = \frac{3GM_0 R^2 \Omega}{10 c^2 r^3} \qquad (65)$$

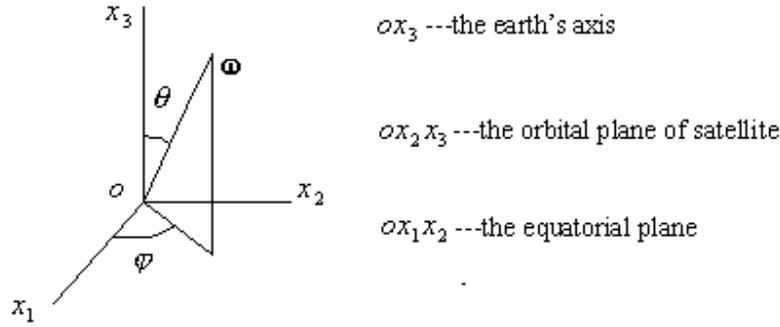

Fig. (1) The definite of $\varphi$ and $q$

$\varphi$ is the angle between the direction of projection of gyroscope angular velocity in $ox_1 x_2$ and $ox_1$ (the normal line of orbital plane), $q$ is the angle between $ox_3$ and the direction of the angular velocity of the gyroscope. Fig. (2) illustrates the relation between $\dot{\psi}_S$ with $\varphi$ in both gravitational theories.

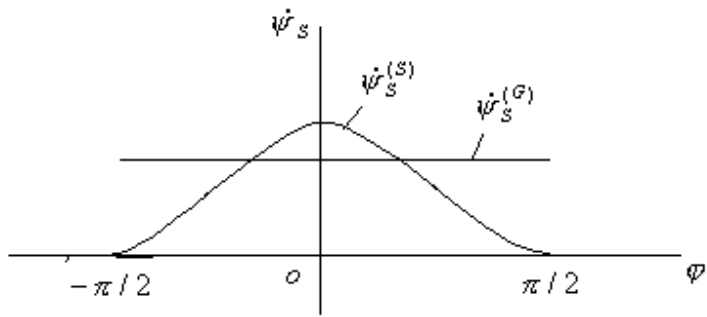

Fig. 2 The relation between $\dot{\psi}_S$ with $\varphi$ in both gravitational theories

If $\varphi \approx \pm 35.3^0$, $\dot{\psi}_S^{(S)} \approx \dot{\psi}_S^{(G)}$, it is then extremely difficult to distinguish which gravitational theory is correct from the experiment's value. The most robust case would be if we set one gyroscope that its revolving direction is in $ox_1x_3$ plane ($\varphi = 0$), and another gyroscope its revolving direction is in $ox_2x_3$ plane ($\varphi = \pi/2$). By comparing the procession rates, we can not only judge which gravitational theory is correct, but also determine the rate of two contributions of **B** and $\wp$ terms.

As we discussed earlier that Eq. (8) is a linear approximation. In the case of weak field, the gravitational radiation effects are weak. We can only observe gravitational radiation effects in the motion of extra dense body, such as a supernova. In this case we must use a nonlinear field equation to substitute Eq. (8). We shall discuss this problem in a separate paper.